\documentclass{article}
\usepackage{ismir,amsmath,cite,url}
\usepackage{graphicx}
\usepackage{color}
\usepackage{array}
\usepackage{amssymb}
\usepackage{booktabs}
\usepackage{enumitem}
\usepackage[hang,flushmargin]{footmisc}

\title{Score-informed syllable segmentation for a cappella singing voice with convolutional neural networks}

\oneauthor
{Jordi Pons$^*$, Rong Gong$^*$ and Xavier Serra}
{Music Technology Group, Universitat Pompeu Fabra, Barcelona\\ {\tt jordi.pons@upf.edu, rong.gong@upf.edu, xavier.serra@upf.edu}\\\tt\small $^*$ contributed equally}

\begin{document}

\maketitle
\begin{abstract}
This paper introduces a new score-informed method for the segmentation of jingju a cappella singing phrase into syllables. The proposed method estimates the most likely sequence of syllable boundaries given the estimated syllable onset detection function (ODF) and its score.
Throughout the paper, we first examine the jingju syllables structure and propose a definition of the term ``syllable onset". Then, we identify which are the challenges that jingju a cappella singing poses. 
Further, we investigate how to improve the syllable ODF estimation with convolutional neural networks (CNNs). We propose a novel CNN architecture that allows to efficiently capture different time-frequency scales for estimating syllable onsets.
In addition, we propose using a score-informed Viterbi algorithm --instead of thresholding the onset function--, because the available musical knowledge we have (the score) can be used to inform the Viterbi algorithm in order to overcome the identified challenges.
The proposed method outperforms the state-of-the-art in syllable segmentation for jingju a cappella singing. We further provide an analysis of the segmentation errors which points possible research directions.
\end{abstract}
\section{Introduction}\label{sec:introduction}

\begin{figure}[ht!]
    \centering
    \includegraphics[width=8.5cm]{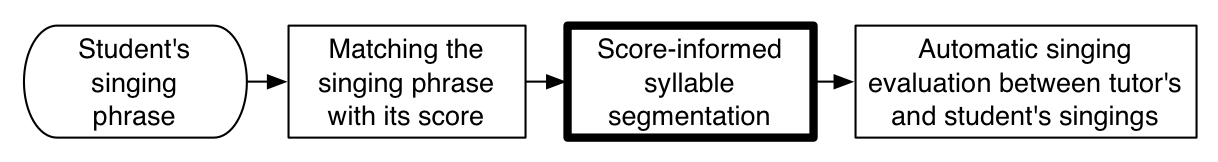}
    \caption{Framework of the entire research project. The module with bold border is addressed in this paper.}
    \label{fig:design_framework}
\end{figure}

The ultimate goal of our research project is to automatically evaluate the jingju a cappella singing of a student in the scenario of jingju singing education -- see \figref{fig:design_framework}. Jingju, a traditional Chinese performing art form also known as Peking or Beijing opera, is extremely demanding in the clear pronunciation and accurate intonation for each syllabic or phonetic singing unit. To this end, during the initial learning stages, students are required to completely imitate tutor's singing. Therefore, the automatic jingju singing evaluation tool we envision is based on this training principle and measures the intonation and pronunciation similarities between the student's and the tutor's singings. Before measuring the similarities, the singing phrase should be automatically segmented into syllabic or phonetic units in order to capture the temporal details. 

\noindent In this paper we tackle the problem of score-informed automatic syllable segmentation for a cappella singing  (bold rectangle in \figref{fig:design_framework}). Jingju music scores, which contain the duration information for each singing syllable, will be a helpful hint for the segmentation method. 

\begin{figure*}
\centering
\includegraphics[width=\textwidth]{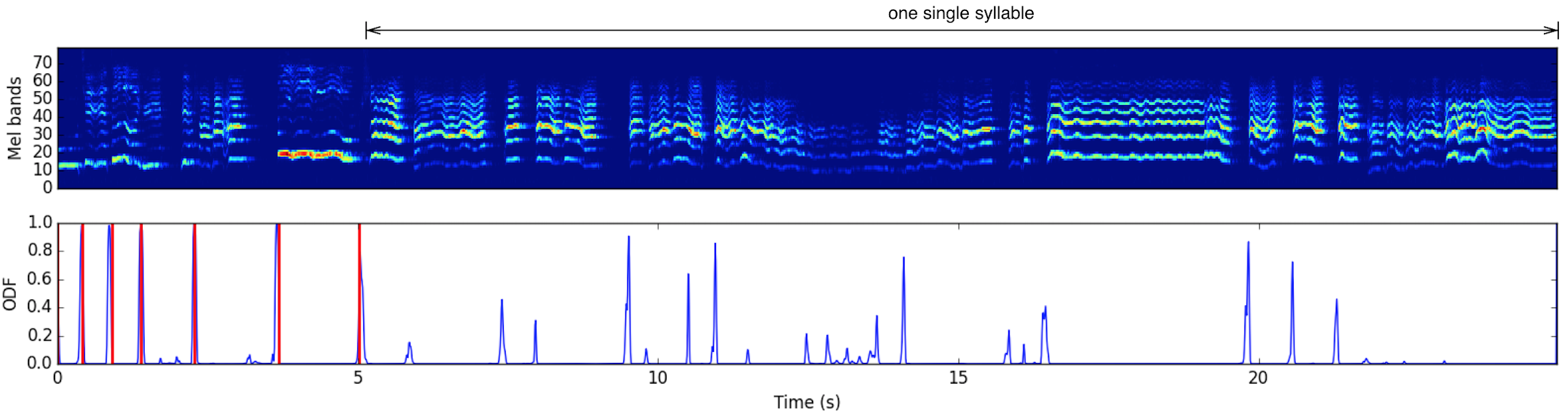}
\caption{A \textit{dan} role-type singing phrase with its last syllable prolongated. \textbf{Top:} log-mel bands spectrogram.\hspace{15mm}
\textbf{Bottom:} syllable ODF estimated with Schl\"{u}ter's method \cite{schluter2014} (blue), and ground truth syllable onsets (red).}
\label{fig:challenge}
\end{figure*}

The syllable segmentation task consists on determining the time positions of the syllable boundaries -- onset and offset. In this research, we consider the onset of the subsequent syllable to be the offset of the current one. Therefore, we treat the segmentation task as an onset detection problem.

Most segmentation methods rely on first estimating an ODF. For example, Klapuri \cite{klapuri1999} utilizes a band-wise processing principle inspired by psychoacoustics. 
He divides the audio signal into 21 non-overlapping bands, then detects onset components in each band ODF and finally combines them to yield onsets. 
Obin \textit{et al.} \cite{obin_syll-o-matic:_2013} obtain a syllable ODF by fusing mel-frequency intensity profiles and voicing profiles.
One shortcoming of these band-wise methods has been already pointed out by Klapuri \cite{klapuri1999}: \textit{``they are unable to deal with strong amplitude modulations"} -- what is very common in singing voice recordings. 

On the other hand, some methods are based on features and supervised learning. For example, Toh \textit{et al.} \cite{Toh2008MultipleFeatureFB} use two Gaussian Mixture Models (GMMs) to classify singing audio frames into onset or non-onset classes. Three timbral features (MFCCs, LPCCs and ERB-bands) are chosen as input. Their results show that this supervised method is superior to many band-wise ODF-based methods.

Neural networks have also been successfully explored for the task of musical onset detection. Eyben \textit{et al.} \cite{eyben2010universal} trained a bidirectional long-short term memory neural network on mel-scale magnitude spectrograms to estimate an ODF. Schl\"{u}ter \textit{et al.} \cite{schluter2014} proposed to use CNNs to estimate the ODF, which defines the current state-of-the-art for onset detection.
The advantage of applying deep learning methods in musical onset detection is that one does not need to design rules or handcraft features to capture the relevant facets for detecting onset/non-onset frames -- which are very difficult to design. Besides, if context (more than one frame) is input into the network, spectro-temporal features can also be learned -- what might be useful to learn slow-transient features defining some onsets.

Thresholding or peak-picking operations are normally executed over the ODF in order to determine the final onsets \cite{klapuri1999,Toh2008MultipleFeatureFB,eyben2010universal,schluter2014}. However, we will argue that these operations are not suitable for selecting jingju singing syllable onsets. Probabilistic models --which allow incorporating \textit{prior} domain knowledge for decision making-- usually result in a performance gain compared to simple thresholding and peak-picking methods. 
For example, B\"{o}ck \textit{et al.} \cite{bock2016joint} tracked the beat/downbeat by using a dynamic bayesian network (DBN) observing a beat ODF estimated by a recurrent neural network (RNN). 
Or Obin \textit{et al.} \cite{obin_syll-o-matic:_2013} applied a segmental Viterbi algorithm over a syllable ODF to detect speech syllable onsets. These two approaches are closely related to ours: former one relates with our work because it uses deep learning to estimate the ODF, and latter one because we also consider using the Viterbi algorithm for estimating the onset candidates. 

This paper introduces a new score-informed method for the segmentation of jingju singing into syllables. We first define what a ``syllable" and ``syllable onset" is in the context of jingju music. By doing so, in section \ref{sec:background} we introduce the challenges we aim to address. The audio dataset we use is described in section \ref{sec:dataset}. Section \ref{sec:approach} explains the CNN architecture used for estimating the syllable ODF, and the Viterbi decoding algorithm that exploits the prior syllable duration information extracted from the score. Evaluation and error analysis are conducted in section \ref{sec:experiments}, and section \ref{sec:conclusions} concludes this work.

\section{Background and Challenges}\label{sec:background}


Jingju singing is the most precise articulated rendition of the spoken Mandarin language. 
Although certain special pronunciations in jingju theatrical language differ from their normal Mandarin pronunciations --due to: firstly, the adoption of certain regional dialects; and secondly, the ease or variety in pronunciation and projection of sound-- the mono-syllabic pronouncing structure of the standard Mandarin doesn't change \cite{wichmann_listening_1991}. 

A syllable/character of jingju singing is composed of three distinct parts in most of the cases: the ``head" (\textit{tou}), the ``belly" (\textit{fu}) and the ``tail" (\textit{wei}). 
The ``head" consists of an initial consonant or semi-vowel -- and a medial vowel, if the syllable includes one. The ``head" is not normally prolonged in its pronunciation except when there is a medial vowel. 
The ``belly" follows the ``head"  and consists of the central vowel and it is normally prolonged. The ``belly" is the most sonorous part of a jingju singing syllable and can be analogous to the nuclei of a speech syllable. The ``tail" is composed of the terminal vowel or consonant \cite{wichmann_listening_1991}. However, there are syllables in jingju singing where ``head" is absent -- only ``belly" is an obligatory element. 
To avoid ambiguity, we define the syllable onset as: \textit{``the start of the initial consonant if the syllable includes one or the start of the central vowel otherwise"}. This definition is slightly different from that agreed by most of the phonological theories \cite{kessler1997syllable}: ``\textit{any consonants that precede the nuclear element (the vowel)}". It is also worth stressing that our notion of singing syllable onset is different from the singing onset defined in Toh \textit{et al.}'s paper \cite{Toh2008MultipleFeatureFB}: ``\textit{the start of a new human-perceived note, taking into account contextual cues}", of which the latter emphasizes on the intonational aspect instead of the phonological aspect of the singing voice.

\subsection{Challenges}\label{sec:challenge}
\figref{fig:challenge} shows an example of a \textit{dan} role-type singing phrase in which the last syllable lasts approximately 20s. This singing method is more common in \textit{dan} singing than in \textit{laosheng} singing. However, both role-types use it as a way of improving artistic expression and showing off their singing skills. These skills include breathing techniques, intonational techniques and dynamic control techniques, among others. The syllable ODF (blue curve in \figref{fig:challenge}) is generated using a CNN model based on Schl\"{u}ter \textit{et al.}'s work \cite{schluter2014}, which is considered the state-of-the-art. This CNN model is trained with the jingju dataset presented in section \ref{sec:dataset}. The resulting syllable ODF is quite robust to pitch variations and continuous dynamic change since no prominent peaks are observed in these regions. However, numerous peaks can be found in the start and end positions of each articulation throughout this long syllable ($\approx$ 20s). Note that thresholding or peak-picking would choose these peaks (false syllable onsets) since they have similar amplitude level than real onsets. For that reason, we propose using a score-informed decoding method as an alternative to naive thresholding.

\begin{figure}[ht!]
\centering
\includegraphics[width=8cm]{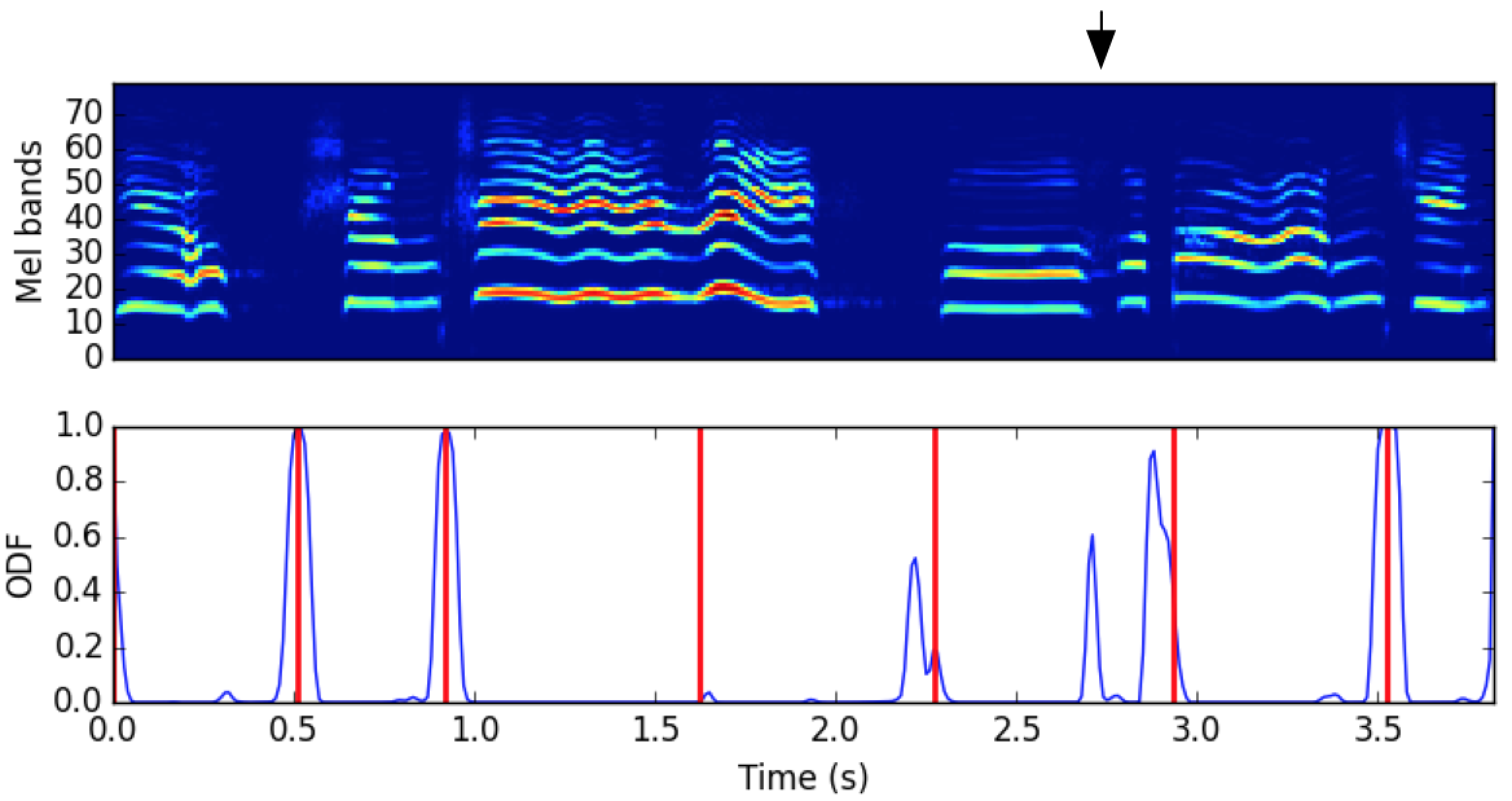}
\caption{Another \textit{dan} role-type singing phrase. A \textit{zu yin} (block sound) is located by the arrow within a normal length syllable. Schl\"{u}ter's syllable ODF (blue curve). Ground truth syllable onset positions (red vertical lines).}
\label{fig:challenge2}
\end{figure}

Stealing breath (\textit{tou qi}) is one of the major methods of taking breath in jingju singing. A breath is stolen when a sound is too long to be delivered in one breath -- and no vocal pauses are desired \cite{wichmann_listening_1991}. However, this is not the only technique which can lead to pauses within a syllable. Another singing technique (\textit{zu yin} -- literal translation: block sound), provokes also pauses without occurring exhalation or inhalation. This kind of pause can be very short in duration and can be easily found in jingju singing syllables, see \figref{fig:challenge2} for an example. These silences can be another source of false positives if thresholding or peak-picking is used. However, these can also be avoided by using an score-informed decoding method.

\textit{A priori} syllable duration information is often easy to obtain from the score and this is an advantage which we exploit to avoid previously described issues. The repertoire of jingju includes around 1400 plays \cite{wichmann_listening_1991}, and most used teaching pieces are transcribed into scores. We use the score to guide the Viterbi decoding throughout the onset detection process.

\section{Dataset} \label{sec:dataset}
The \textit{a cappella} singing audio dataset\footnote{https://goo.gl/y0P7BL} used for this study focuses on two most important jingju role-types \cite{repetto_creating_2014}: \textit{dan} (female) and \textit{laosheng} (old man). It contains 39 interpretations of 31 unique arias sung by 11 jingju singers. Audio is sampled at 44.1 kHz and is is pre-segmented into phrases. The syllable onset ground truth is manually annotated in Praat \cite{boersma_praat_2001} -- with 291 phrases and 2641 syllables (including padding written-characters \cite{wichmann_listening_1991}). 
Two Mandarin native speakers and a jingju musicologist have been devoted to this annotation work. Syllable durations are manually transcribed from music scores considering as unit duration the quarter note. The whole dataset is randomly split into training, validation and test sets (60\%, 20\% and 20\%, respectively). The percentage of presence of each role-type in a split is kept constant throughout sets.

In order to highlight the differences between jingju music and popular western music, we compare the vowel's duration between Kruspe's dataset \cite{kruspe14} (a cappella singing of commercial pop songs) and ours. In Kruspe's dataset, the standard deviation duration of voiced phonemes is of 0.31s, whereas this duration is more than doubled in our dataset: 0.75s. 
Since voiced phonemes are the main component of a syllable, it thus becomes clear that jingju singing syllable durations show huge variations. Therefore, it is impossible to model syllable durations with a single distribution. For that reason the proposed model must be able to handle different syllable lengths depending on the case. To this end, syllable durations extracted from scores will be a valuable information.

\section{Approach}\label{sec:approach}
A score-informed syllable boundary detection approach is explained in this section. We first propose some improvements to the current state-of-the-art CNN onset detection model proposed by Schl\"{u}ter \textit{et al.} The syllable ODF output from the CNN serves as observation probability for the syllable boundary decoding process. Then, an \textit{a priori} syllable duration model based on the score is proposed. This model guides the Viterbi algorithm by informing it in which time-positions is likely to occur a syllable boundary.
Therefore, the syllable boundaries sequence is decoded by taking advantage of a score-informed Viterbi algorithm.

\subsection{CNN syllable onset detection function}\label{sec:cnn_architecture}

Studied CNNs are inspired by Schl\"{u}ter \textit{et al.}'s \cite{schluter2014} and Pons \textit{et al.}'s work \cite{pons2017designing,pons2017timbre}. Schl\"{u}ter \textit{et al.} have shown that CNNs fed with spectrograms can achieve state-of-the-art performance for the onset detection task \cite{schluter2014}. On the other hand, Pons \textit{et al.} \cite{pons2017designing} have recently proposed a novel design strategy for spectrograms-based CNNs. They propose using different filter shapes in the first layer so that local stationarities in spectrograms (present at different time/frequency-scales) can be efficiently captured.

Explored models are based on Schl\"{u}ter \textit{et al.}'s architecture \cite{schluter2014}, which consists of: two convolutional layers, and a dense layer of 256 units connected to an output softmax layer -- with two output units standing for onset and non-onset.
First CNN layer uses 20x $3{\times}7$ filters\footnote{First number denote the number of filters (\textit{ie.} 20x). Second and third respectively denote the frequency and temporal size of the filter (\textit{ie.} $3{\times}7$).} and is followed by a $3{\times}1$ max pool layer. Second CNN layer has 20x $3{\times}3$ filters and is followed by a $3{\times}1$ max-pool layer. 
Input is set to be a log-mel spectrogram\footnote{Original Schl\"{u}ter \textit{et al.}'s model inputs three channels with different resolution spectrograms to the network. However, preliminary results showed that a single channel was performing better than three.} of size $80{\times}21$ -- note that the network takes a decision for every frame given its context: $\pm$10ms, 21 frames in total.

\begin{figure}[h]
	\centering
	\includegraphics[width=8cm]{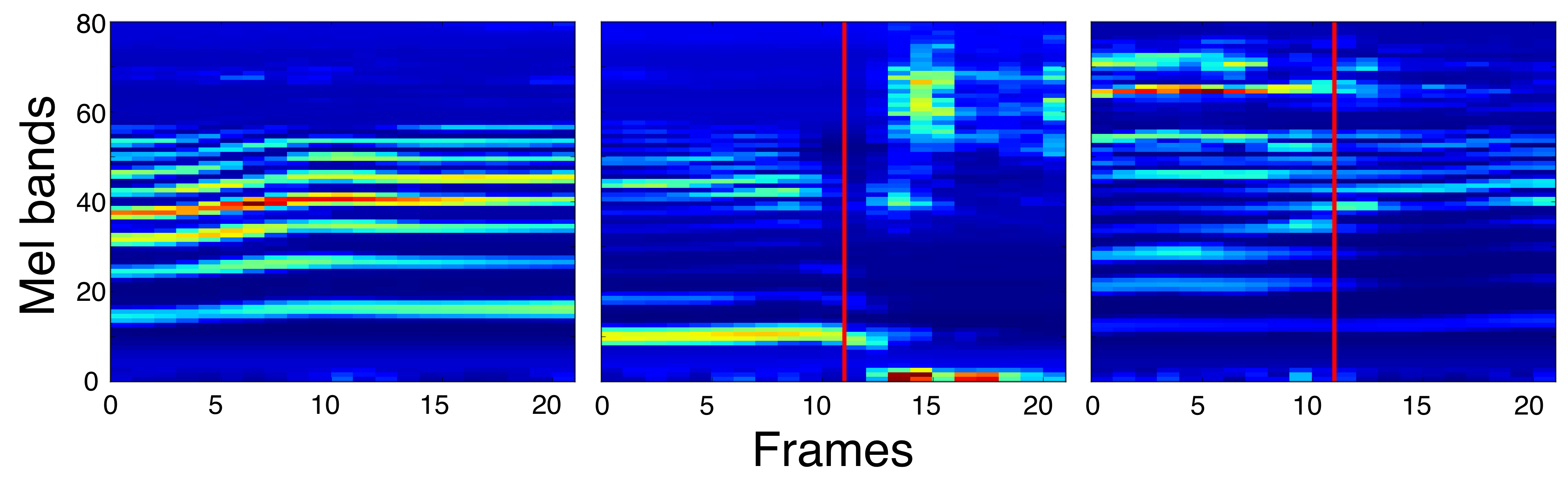}
	\caption{Log-mel spectrograms (80 bands, 21 frames). \textit{left}: Non-onset spectrogram. \textit{middle} and \textit{right}: Onset spectrograms. Red line denotes the onset frame.}
	\label{fig:transient}
\end{figure}

Proposed architectures follow Pons \textit{et al.} \cite{pons2017designing,pons2017timbre} design strategy and several filter shapes in the first layer are used. This approach allows to efficiently capture different time-frequency scales, what might be interesting for the task at hand because onsets can exhibit different time-frequency patterns. For example, Figure \ref{fig:transient} depicts two onsets: \textit{middle} onset is expressed as an abrupt time-frequency change corresponding to an onset starting with a consonant, and \textit{right} onset is expressed as a gradual timbral change that corresponds to an onset starting with a central vowel. Note that these examples are representative of the proposed definition of syllable onset. Moreover, using different filter shapes in the first layer promotes a much richer representation out of the first layer. In the following, several CNNs are designed to efficiently capture the relevant time-frequency contexts and scales for onset detection.

Schl\"{u}ter \textit{et al.}'s architecture and proposed ones only differ in the way the first convolutional layer and its following max-pool layer are set up. Input and remaining layers are kept intact -- unless it is explicitly stated:
 
\noindent  \textbf{$\rightarrow$} \underline{Temporal architecture}.
Note that the small filters proposed by Schl\"uter \textit{et al.} might have difficulties on learning the longer time-scales required to model the gradual changes that some onsets express -- see Figure \ref{fig:transient}, \textit{right}.
Several filter shapes are designed to efficiently capture different and longer time-scales \cite{pons2017designing} than Schl\"uter \textit{et al.}: 

\hspace{1mm}$\cdot$ 12x $1{\times}7$, 6x $3{\times}7$ and 3x $5{\times}7$

\hspace{1mm}$\cdot$ 12x $1{\times}12$, 6x $3{\times}12$ and 3x $5{\times}12$

\noindent These CNNs are followed by a $3{\times}5$ max-pool layer. 


\noindent \textbf{$\rightarrow$} \underline{Timbral architecture}. Since our syllable onset definition considers phonological aspects, we consider the timbre to be an important feature for our task.
Filter shapes are designed to learn timbral representations, note that these filters span throughout the frequency domain \cite{pons2017timbre}:

\hspace{1mm}$\cdot$ 12x $50{\times}1$, 6x $50{\times}5$ and 3x $50{\times}10$

\hspace{1mm}$\cdot$ 12x $70{\times}1$, 6x $70{\times}5$ and 3x $70{\times}10$

\noindent These CNNs are followed by a $5{\times}3$ max-pool layer.


We also explore combining temporal and timbral architectures. We propose a late-fusion approach where we multiply the estimated output probabilities from temporal and timbral (independently trained) architectures\footnote{Using temporal and timbral filters in first layer yield to worse results.}.

In addition, we also explore increasing the number of filters in the second layer from 20x to 32x. 

For concatenating several feature maps resulting of CNNs with different filter shapes, it is required to use zero padding. We apply \textit{same} padding in the first CNN layer, so that all resulting feature maps have the same length and these can be concatenated. 
STFT was performed using a 25ms window (2048 samples with zero-padding) with a hop size of 10ms. The 80 log-mel bands energies are calculated on frequencies between 0Hz and 11000Hz and spectrograms are standardized to have zero mean and unit variance.
We use L2 weight decay regularization, ELU activation functions \cite{ClevertElu} and 30\% dropout for each layer. The model parameters are learned with mini-batch training (batch size 128) using the ADAM update rule \cite{kingma2014adam} and early stopping -- if validation loss (categorical cross-entropy) was not decreasing after 10 epochs. In order to allow a fair comparison between Schl\"{u}ter's and Pons' architectures, the original Schl\"{u}ter architecture is modified to meet above described hyper-parameters.

Finally, syllable ODFs estimated with CNNs are smoothed by convolving those with a 5 frames Hanning window \cite{schluter2014} -- since estimated syllable ODFs are typically very spiky.

\subsection{A priori duration model}
\label{sec:aprioriDurDist}
\begin{figure}[h!]
\centering
\includegraphics[width=8cm]{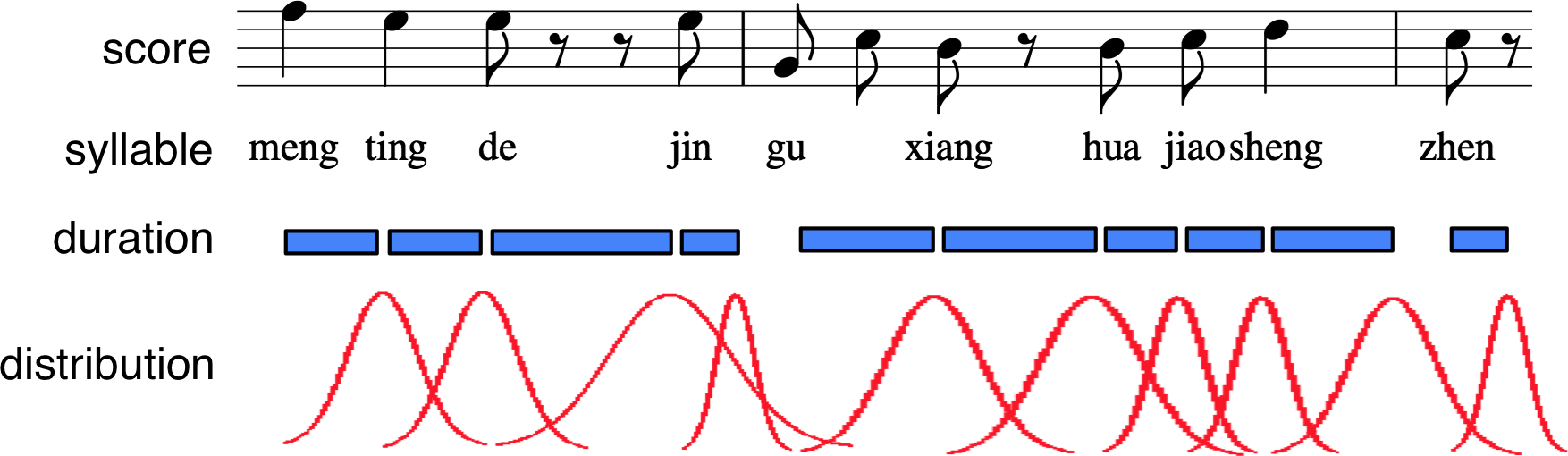}
\caption{\textit{A priori} relative duration distributions (bottom) of the syllables of a singing phrase.}
\label{fig:durationDistrScheme}
\end{figure}

The \textit{a priori} duration model is shaped with a Gaussian function $\mathcal{N} (x; {\mu}_l, \sigma_l^2)$ whose mean ${\mu}_l$ represents the $l$-th relative syllable duration -- according to the score. Its standard deviation $\sigma_l$ is proportional to $\mu_l$: $\sigma_l=\gamma \mu_l$ and $\gamma$ is heuristically set to 0.35. 
\begin{equation}
\label{eq:sylDurDist}
\mathcal{N} (x ; \mu_l, \sigma_l^2) = \frac{1}{\sqrt{2 \pi} \sigma_l} \exp \left(-\frac{(x-\mu_l)^2}{2\sigma_l^2} \right).
\end{equation}
Figure \ref{fig:durationDistrScheme} provides an intuitive example of how the \textit{a priori} duration model works. Observe that it provides the prior likelihood of an onset to occur according to the duration in the score.

The relative duration of each note is measured considering a quarter note length as a unit, so an eighth note has a duration of 0.5. We only keep the relative duration and discard the tempo information of the score. By normalizing the summation of the notes' relative durations to the incoming audio recording's duration, we obtain the absolute notes' durations. Then, a sequence of syllable absolute durations $M=\mu_1 \mu_2 \cdots \mu_L$ is deduced by summing the notes' absolute durations corresponded with each syllable, where $L$ is the total syllable number of the score. The a priori duration model distributions will be incorporated into the Viterbi algorithm as state transition probabilities to inform the algorithm where syllable boundary are likely to occur.

\subsection{Decoding of the syllable boundaries}\label{sec:decoding}
To decode the syllable boundaries, we construct a hidden markov model characterized by the following:
\begin{enumerate}[noitemsep]
    \item The hidden state space is a set of $N$ candidate onset positions $S_1, S_2, \cdots, S_N$ discretized by the hop size, where $S_{N}$ is the offset position of the last syllable. 
    \item The state transition probability at the time instant $l$ associated with state changes is defined by \textit{a priori} duration distribution $\mathcal{N} (d_{ij} ; \mu_l, \sigma_l^2)$, where $d_{ij}$ is the time distance between states $S_i$ and $S_j$ ($j>i$). The length of the decoded state sequence is equal to the total syllable number $L$ written in the score. 
    \item The observation probability for the state $S_j$ is represented by its corresponding value in the syllable ODF $p$, which is denoted as $p_j$.
\end{enumerate} 
The goal is to find the best onset position state sequence $Q={q_1 q_2 \cdots q_L}$ for a given \textit{a priori} duration sequence $M$, where $q_i$ denotes the onset of the $i+1$th decoding syllable. $q_0$ and $q_L$ are fixed as $S_1$ and $S_N$ as we expect that the onset of the first syllable to be located at the beginning of the incoming audio and the offset of the last syllable is located at the ending of the audio. One can fulfill this assumption by truncating the silences at the beginning and at the end of the incoming audio. According to the logarithmic form of Viterbi algorithm \cite{rabiner_tutorial_1989}, we define:
\begin{equation*}
\delta_l(i)= \max_{q_1,q_2,\cdots,q_l}{\log P[q_1 q_2 \cdots q_l,\, \mu_1 \mu_2 \cdots \mu_l]}
\end{equation*}
with the initial step as follows:
\begin{align*}
\delta_1(i) &=\log(\mathcal{N} (d_{1i} ; \mu_1, \sigma_1^2))+\log(p_i) \\
\psi_1(i) &= S_1
\end{align*}
with the following recursive step:
\begin{align*}
\delta_l (j) &= \max_{1 \leqslant i < j} [\delta_{l-1} (i) + \log(\mathcal{N} (d_{ij} ; \mu_l, \sigma_l^2))] +\log(p_j) \\
\psi_l (j) &= \arg\max_{1 \leqslant i < j} [\delta_{l-1} (i) + \log(\mathcal{N} (d_{ij} ; \mu_l, \sigma_l^2))]
\end{align*}
and the following termination step:
\begin{align*}
\log P^* &= \max_{1 \leqslant i < N} [\delta_{L-1} (i) + \log(\mathcal{N} (d_{i N} ; \mu_L, \sigma_L^2))] \\
q_{L}^* &= \arg\max_{1 \leqslant i < N} [\delta_{L-1} (i) + \log(\mathcal{N} (d_{i N} ; \mu_L, \sigma_L^2))]
\end{align*}

\begin{figure}[h!]
\centering
\includegraphics[width=6.9cm]{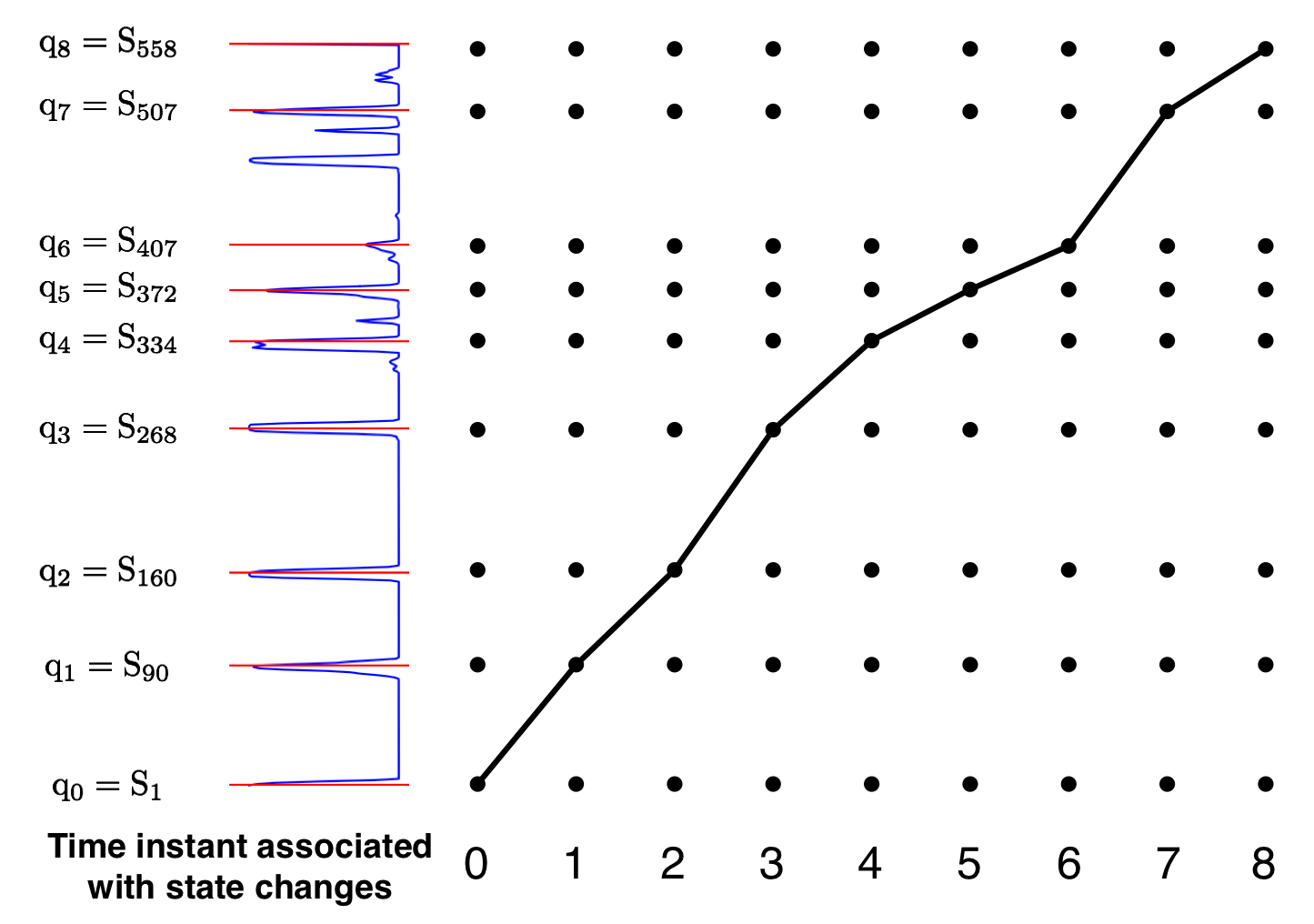}
\caption{Illustration of an example of the best boundary position decoding path (black) with $N=558$ and $L=8$. The intermediate states are omitted in the vertical direction for a clear visualization. Blue curve: syllable ODF; red horizontal lines: decoded boundary positions.}
\label{fig:viterbiDecoding}
\end{figure}

Finally, the best offset position state sequence $Q$ is obtained by the backtracking step. An example of the best boundary position decoding path is showed in figure \ref{fig:viterbiDecoding}.

\section{Experiments and results}\label{sec:experiments}
\subsection{Performance metrics}
The syllable segmentation task consists on determining the time positions of syllable boundaries.
The proposed evaluation consists in comparing the detected syllable onsets/offsets to their reference ones. We report F-measure results in this paper. 
The definition of a correct segmented syllable is borrowed from the note transcription literature \cite{molina_evaluation_2014}. For syllable onsets, we choose an evaluation tolerance of $\pm\tau$ ms. For offsets, we choose an evaluation tolerance of either \textit{(a)} $\pm$20\% of the syllable's duration annotation, or \textit{(b)} $\pm\tau$ ms -- whichever is larger. If both the onset and the offset of a syllable lie within the tolerance of their annotated counterparts and the syllable is correctly labeled, we consider that it's correctly segmented. We report the results for a tolerance of $\tau=0.05$ (seconds).

\subsection{Results and discussion}\label{sec:results}

Two state-of-the-art methods are set as baselines: Obin \textit{et al.} \cite{obin_syll-o-matic:_2013} as a traditional approach, and Schl\"{u}ter \textit{et al.} \cite{schluter2014} as a deep learning method -- both already introduced. We explore Pons \textit{et al.}'s CNNs design strategy \cite{pons2017designing,pons2017timbre} as a way to improve our results. Code is available online\footnote{https://github.com/ronggong/jingjuSyllabicSegmentaion/tree/v0.1.0}.
All syllable ODFs are decoded by using the same approach (described in section \ref{sec:decoding}). F-measure results are reported in Table \ref{table:performanceResults}. The evaluation can not be performed by using peak-picking because the syllabic label can not be attached. However, in section \ref{sec:background} we already discussed the potential problems of peak-picking which are explicitly addressed with the proposed score-informed method.
\vspace{-4mm}
\begin{table}[!ht]
	\centering
	\caption{Syllable segmentation results using different methods for estimating the syllable ODF. Numbers in parenthesis indicate the \textit{\#}filters in the second CNN layer.}
	\label{table:performanceResults}
	\begin{tabular}{lcc}
		\toprule
		\textbf{Syllable ODFs }               & \textbf{\textit{\#}params} & \textbf{F-measure (\%)} \\
		\midrule
		Late-fusion \textit{(32)}    & -     & \textbf{86.37} \\
		Temporal \textit{(32)}       & 210,403  & 83.28 \\
		Timbral \textit{(32)}        & 185,420  & 84.83 \\\midrule
		Late-fusion \textit{(20)}    & -     & 84.05 \\
		Temporal \textit{(20)}       & 132,127  & 83.86 \\
		Timbral \textit{(20)}        & 119,432  & 82.89 \\\midrule
		Schl\"{u}ter \textit{et al.}  \cite{schluter2014}      &535,290        & 81.93 \\
		Obin \textit{et al.} \cite{obin_syll-o-matic:_2013}             & -          & 40.61 \\
		\bottomrule
	\end{tabular}
\end{table}

\begin{figure}[h!]
	\centering
	\includegraphics[width=8cm]{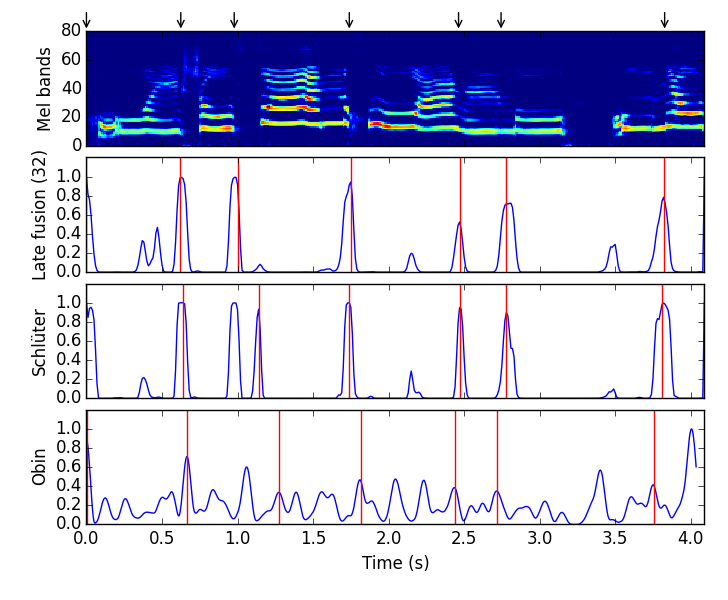}
	\caption{Three syllable ODFs. Red vertical line: decoded syllable onset time positions. Black arrows: ground truth.}
	\label{fig:odf}
\end{figure}

Results show that proposed architectures improve the state-of-the-art in all cases. These results support the idea that using different filter shapes in the first CNN layer is beneficial. Moreover, observe that \textit{\#}params is reduced at least to the half -- interpret \textit{\#}params (number of parameters of the CNN) as a measure for the representational capacity of the network. This denotes how efficient can be CNNs if these are designed to capture the relevant features for the task at hand. Also note that if the \textit{\#}params of a network is reduced, the over-fitting risk is reduced as well.

We also observe that late-fusing the predictions helps improving the results. The interesting idea behind this approach is that individual networks still need to be able to solve the task by their own means -- and we propose fusing models that are tailored towards learning complementary time-frequency scales: temporal and timbral architectures.

We also see a big performance gap between Obin \textit{et al.} and CNN-based methods. For example in \figref{fig:odf} numerous noisy peaks can be observed for Obin \textit{et al.}'s method, what is significantly decreasing its performance. This result denotes the difficulty of designing rules or handcrafting features for estimating the syllable ODF.

\subsection{Error analysis}\label{sec:error}

We conduct an error analysis to study the causes of the segmentation errors produced by our best performing method, what points out future research directions for further improvement.
We study falsely detected onsets in the test dataset by comparatively observing the detected syllable onset positions, the ground truth positions, the spectrogram and the corresponding scores. 
Errors out of 0.05s tolerance are analyzed and 70 syllable onsets out of 512 are identified as false detections. 
Errors are then classified according to their causes. Two main error classes are found: score-performance mismatching and ambiguous syllable transitions.

The majority of the errors (71.42\%) are caused by performance deviations from the score: syllable insertion, syllable deletion or performing different syllable durations than the score. 
The proposed syllable boundary decoding algorithm is not able to preserve the correctness when the difference between the score and its performance becomes too large. Furthermore, since the length of the decoded state sequence of the algorithm must be equal to the number of syllables in the score, singing syllable insertions and deletions not expressed in the score can not be handled properly. One possible solution is to incorporate more domain knowledge of jingju singing into the decoding process -- such as considering the probability of a padding-character to be inserted/deleted, or the expectation of prolonging the last syllable in a phrase. 

The second source of errors include ambiguous syllable transitions (15.61\%) -- such as transitions from vowel to vowel, from vowel or to semi-vowel, etc.
These errors are very difficult to be corrected because no prominent spectral changes can be discerned within such transitions \cite{odetteSpeechSeg}. In future investigations we shall detect ``onset regions" rather than ``onset time positions" since this kind of syllable transitions usually manifest themselves as gradual spectral changes.

\section{Conclusions}\label{sec:conclusions}

This paper introduces a new score-informed method for
the segmentation of jingju singing into syllables. Two main contributions are presented in this paper: \textit{(i)} improvements for estimating the syllable ODF with CNNs, and \textit{(ii)} a method for incorporating score information into Viterbi's algorithm for estimating syllable boundaries.

The improvements to the CNN architecture consisted on using different filter shapes in the first layer and late-fusing the predictions of two models -- designed to learn complementary representations (temporal and timbral). By doing so, we increased the expressiveness of the first layer and enabled the networks to efficiently capture different time-frequency scales useful for detecting syllable onsets. Proposed models, with many filter shapes in the first layer, have proven to be more effective than the state-of-the-art model based in a single filter shape in the first layer.

Moreover, we proposed an \textit{a priori} duration model that describes the probability of a syllable boundary given the score. The likelihood of a syllable boundary is shaped with the \textit{a priori} duration model and incorporated into Viterbi's algorithm as state transition probabilities -- this being the core of the proposed score-informed Viterbi algorithm.

We validated the proposed method on a jingju a cappella singing dataset, which achieved better performance than the state-of-the-art. Although the proposed ameliorations helped improving our results, the proposed method has not yet solved the task. To this end, we plan to investigate incorporating more domain knowledge into the decoding process, and to further improve the ODF with RNNs -- that has proven to be very useful for similar tasks \cite{bock2016joint,eyben2010universal}. 

\section{Acknowledgments}
We are grateful for the GPUs donated by NVidia. This work is partially supported by the Maria de Maeztu Programme (MDM-2015-0502) and the European Research Council under the European Union's Seventh Framework Program, as part of the CompMusic project (ERC grant agreement 267583).

\bibliography{ISMIRtemplate}

\begin{thebibliography}{10}

\bibitem{bock2016joint}
Sebastian B{\"o}ck, Florian Krebs, and Gerhard Widmer.
\newblock Joint beat and downbeat tracking with recurrent neural networks.
\newblock In {\em ISMIR}, New York City, USA, 2016.

\bibitem{boersma_praat_2001}
Paul Boersma.
\newblock Praat, a system for doing phonetics by computer.
\newblock {\em Glot International}, 5(9/10):341--345, 2001.

\bibitem{ClevertElu}
Djork{-}Arn{\'{e}} Clevert, Thomas Unterthiner, and Sepp Hochreiter.
\newblock Fast and accurate deep network learning by exponential linear units
  (elus).
\newblock {\em CoRR}, abs/1511.07289, 2015.

\bibitem{eyben2010universal}
Florian Eyben, Sebastian B{\"o}ck, Bj{\"o}rn~W Schuller, and Alex Graves.
\newblock Universal onset detection with bidirectional long short-term memory
  neural networks.
\newblock In {\em ISMIR}, Utrecht, Netherlands, 2010.

\bibitem{kessler1997syllable}
Brett Kessler and Rebecca Treiman.
\newblock Syllable structure and the distribution of phonemes in english
  syllables.
\newblock {\em Journal of Memory and language}, 37(3):295--311, 1997.

\bibitem{kingma2014adam}
D.~Kingma and J.~Ba.
\newblock Adam: A method for stochastic optimization.
\newblock {\em arXiv:1412.6980}, 2014.

\bibitem{klapuri1999}
A.~Klapuri.
\newblock Sound onset detection by applying psychoacoustic knowledge.
\newblock In {\em ICASSP}, Phoenix, USA, 1999.

\bibitem{kruspe14}
Anna~M. Kruspe.
\newblock Keyword spotting in a-capella singing.
\newblock In {\em ISMIR}, Taipei, Taiwan, 2014.

\bibitem{molina_evaluation_2014}
Emilio Molina, Ana~M. Barbancho, Lorenzo~J. Tardón, and Isabel Barbancho.
\newblock Evaluation {Framework} for {Automatic} {Singing} {Transcription}.
\newblock In {\em {ISMIR}}, Taipei, Taiwan, 2014.

\bibitem{obin_syll-o-matic:_2013}
N.~Obin, F.~Lamare, and A.~Roebel.
\newblock Syll-{O}-{Matic}: {An} adaptive time-frequency representation for the
  automatic segmentation of speech into syllables.
\newblock In {\em {ICASSP}}, Vancouver, Canada, 2013.

\bibitem{pons2017designing}
Jordi Pons and Xavier Serra.
\newblock Designing efficient architectures for modeling temporal features with
  convolutional neural networks.
\newblock In {\em ICASSP}, New orleans, USA, 2017.

\bibitem{pons2017timbre}
Jordi Pons, Olga Slizovskaia, Rong Gong, Emilia G{\'o}mez, and Xavier Serra.
\newblock Timbre analysis of music audio signals with convolutional neural
  networks.
\newblock {\em arxiv:1703.06697}, 2017.

\bibitem{rabiner_tutorial_1989}
L.~Rabiner.
\newblock A tutorial on hidden {Markov} models and selected applications in
  speech recognition.
\newblock {\em Proceedings of the IEEE}, 77(2):257--286, February 1989.

\bibitem{repetto_creating_2014}
Rafael~Caro Repetto and Xavier Serra.
\newblock Creating a {Corpus} of {Jingju} ({Beijing} {Opera}) {Music} and
  {Possibilities} for {Melodic} {Analysis}.
\newblock In {\em ISMIR}, Taipei, Taiwan, 2014.

\bibitem{odetteSpeechSeg}
Odette Scharenborg, Vincent Wan, and Mirjam Ernestus.
\newblock Unsupervised speech segmentation: An analysis of the hypothesized
  phone boundaries.
\newblock {\em The Journal of the Acoustical Society of America},
  127(2):1084--1095, 2010.

\bibitem{schluter2014}
J.~Schl{\"u}ter and S.~B{\"o}ck.
\newblock Improved musical onset detection with convolutional neural networks.
\newblock In {\em ICASSP}, Florence, Italy, 2014.

\bibitem{Toh2008MultipleFeatureFB}
Chee-Chuan Toh, Bingjun Zhang, and Ye~Wang.
\newblock Multiple-feature fusion based onset detection for solo singing voice.
\newblock In {\em ISMIR}, Philadelphia, USA, 2008.

\bibitem{wichmann_listening_1991}
Elizabeth Wichmann.
\newblock {\em Listening to {Theatre}: {The} {Aural} {Dimension} of {Beijing}
  {Opera}}.
\newblock University of Hawaii Press, 1991.

\end{thebibliography}

%
%
%
%

\end{document}